\documentclass{article}
\usepackage{frascatiphys,here,graphicx,subfigure,amssymb}

\begin{document}

\title{MILESTONES OF THE PHENIX EXPERIMENT AT RHIC}
\author{M. Csan\'ad for the PHENIX Collaboration    \\
{\em E\"otv\"os University, Budapest 1117, P\'azm\'any P\'eter s.
1/A, Hungary}}

\maketitle
\baselineskip=11.6pt
\begin{abstract}
The latest PHENIX results for particle production are presented in
this paper. A suppression of the yield of high $p_t$ (transverse
momentum) hadrons in central Au+Au collisions is found. In
contrast, direct photons are not suppressed in central Au+Au
collisions and no suppression of high $p_t$ particles can be seen
in d+Au collisions. This leads to the conclusion that the dense
medium formed in central Au+Au collisions is responsible for the
suppression. It is as well found, that the properties of this
medium are similar to the one of a liquid. Further measurements
provide information about the chiral dynamics of the system.
\end{abstract}

\baselineskip=14pt

\section{Introduction}

Ultra-relativistic collisions, so called ``Little Bangs'' of
almost fully ionized Au atoms are observed at the four experiments
(BRAHMS, PHENIX, PHOBOS and STAR) of the Relativistic Heavy Ion
Collider (RHIC) of the Brookhaven National Laboratory, New York.
The aim of these experiments is to create new forms of matter that
existed in Nature a few microseconds after the Big Bang, the
creation of our Universe.

A consistent picture emerged after the first three years of
running the RHIC experiment: quarks indeed become deconfined, but
also behave collectively, hence this hot matter acts like a
liquid~\cite{Adcox:2004mh}, not like an ideal gas theorists had
anticipated when defining the term QGP. The situation is similar
to as if prisoners (quarks and gluons confined in hadrons) have
broken out of their cells at nearly the same time, but they find
themselves on the crowded jail-yard coupled with all the other
escapees. This strong coupling is exactly what happens in a
liquid~\cite{Riordan:2006df}.

\section{High pt suppression}

\begin{figure}
  \begin{center}
  \includegraphics[width=0.49\linewidth]{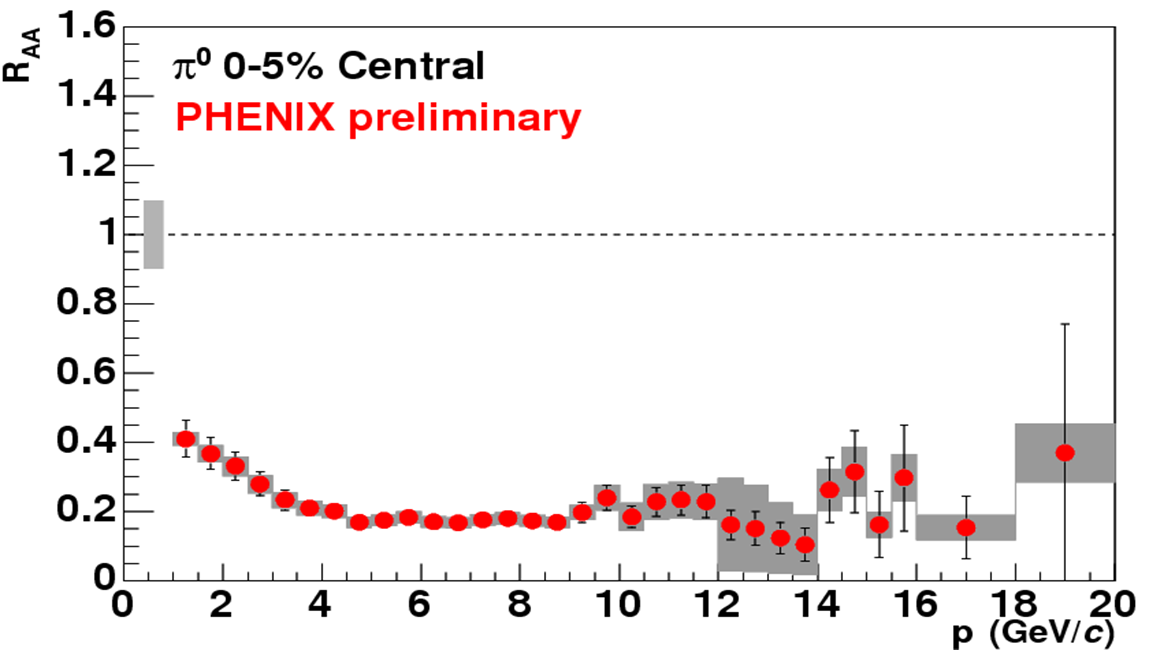}
  \includegraphics[width=0.49\linewidth]{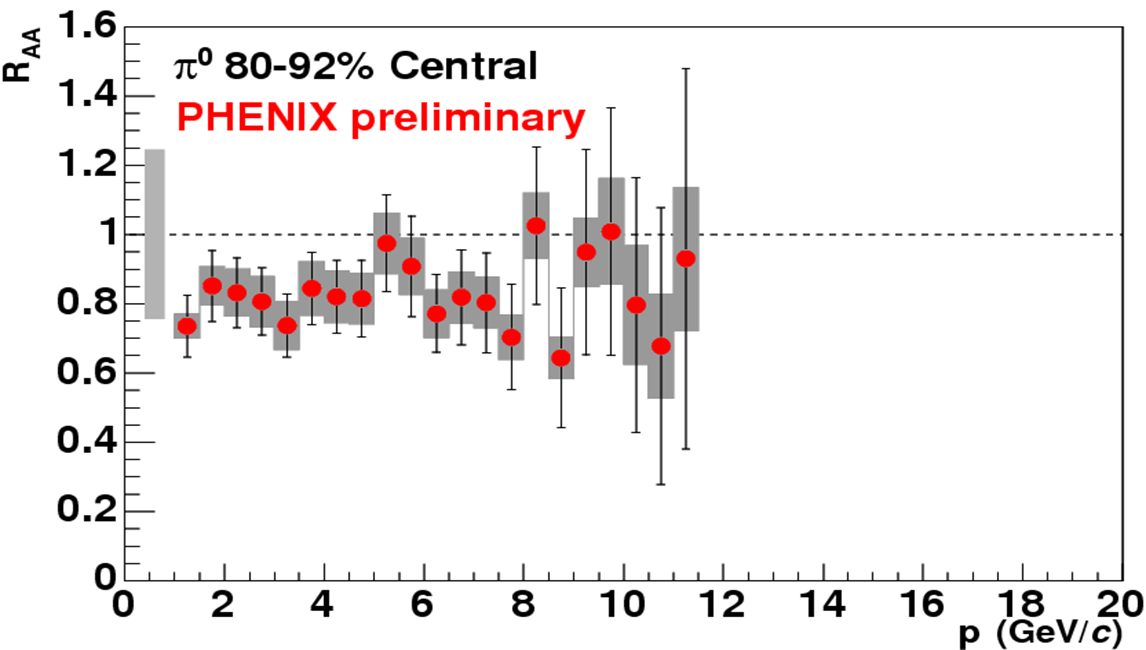}\\
  \includegraphics[width=0.49\linewidth]{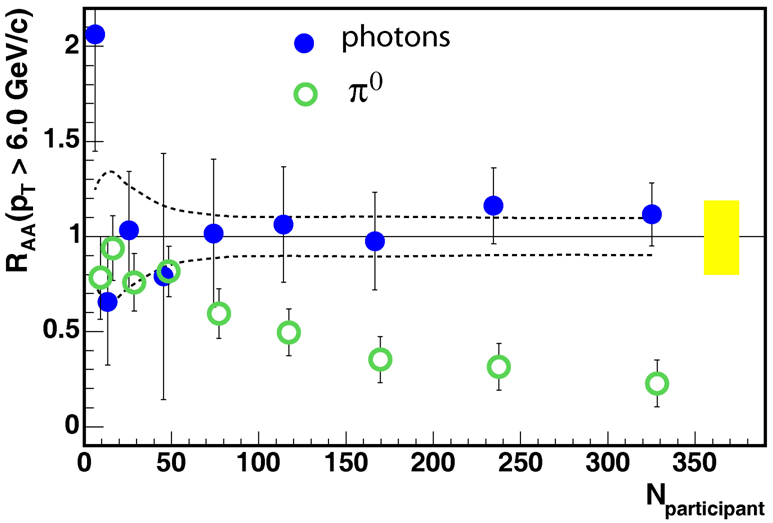}
  \includegraphics[width=0.49\linewidth]{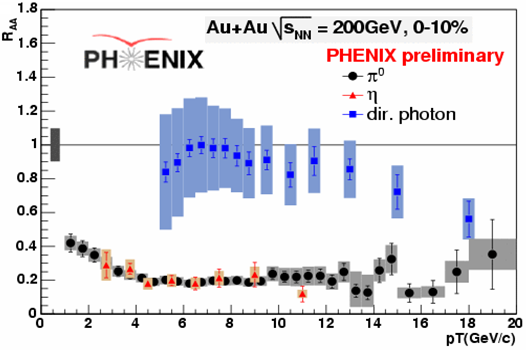}
  \end{center}
  \caption{
Nuclear modification factor $R_{\rm AA}$ for $\pi_0$, $\eta$ and
photon yields in Au+Au collisions as a function of $p_t$ for
different centralities (different number of participants). The
shaded error band around unity indicate systematic errors.}
\label{f:raa}
\end{figure}

High transverse momentum particles resulting from hard scatterings
between incident partons have become one of the most effective
tools for probing the properties of the medium created in
ultra-relativistic heavy ion collisions at RHIC. Nuclear
modification factor, defined as
\begin{equation}
R_{\rm AA}(p_t) \equiv {\textnormal{Yield\ \ in\ \ Au+Au\ \
events} \over \textnormal{Scaled Yield\ \ in\ \  p+p\ \ events}},
\end{equation}
was measured in central and preripheral Au+Au collisions at the
four RHIC
experiments~\cite{Adcox:2001jp,Adcox:2002pe,Adams:2003kv,Adler:2003qi,Adler:2003au,Arsene:2003yk,Back:2003qr,Adler:2006bw}.
The measurements show a high transverse momentum hadron
suppression in central Au+Au collisions compared to (appropriately
scaled) p+p collisions, while there is no such suppression in
peripheral Au+Au or d+Au collisions
\cite{Adler:2003ii,Adams:2003im,Back:2003ns}, as shown in the
upper plots of Fig.~\ref{f:raa}. This shows that the suppression
is not due to modification of parton distributions in the
colliding nuclei.

The nuclear modification factor has been measured for several
hadron species at highest $p_t$: for $\pi_0$, and most recently
$\eta$ mesons\cite{Adler:2006hu}, as shown in the lower plots of
Fig.~\ref{f:raa}. This confirms the above evidence for a dense and
strongly interacting matter. On the other hand, direct photon
measurements, which require tight control of experimental
systematics over several orders of magnitude, show that the high
$p_t$ photons in Au+Au collisions are not
suppressed~\cite{Adler:2005ig} and, thus, provide final
confirmation that hard scattering processes occur at rates
expected from point-like processes. This observation makes
definitive the conclusion that the suppression of high-$p_t$
hadron production in Au+Au collisions is a final-state effect.

\section{The perfect fluid of quarks}

One of the most important results of RHIC is the relatively strong
second harmonic moment of the transverse momentum distribution,
referred to as the elliptic flow. The elliptic flow is an
experimentally measurable observable and is defined as the
azimuthal anisotropy or second Fourier-coefficient of the
single-particle momentum distribution $N_1(p)$. The $n^{\rm th}$
Fourier-coefficient is defined as:
\begin{equation}
v_n = \frac{\int_0^{2 \pi} N_1(p) \cos(n\varphi) d\varphi}
           {\int_0^{2 \pi} N_1(p) d\varphi},
\end{equation}
$\varphi$ being the azimuthal (perpendicular to the beam) axis of
momentum $p$ with respect to the reaction plane. This formula
returns the elliptic flow $v_2$ for $n=2$.

Measurements of the elliptic flow by the PHENIX, PHOBOS and STAR
collaborations (see
refs.~\cite{Back:2004zg,Back:2004mh,Adler:2003kt,Adams:2004bi,Adler:2001nb,Sorensen:2003wi})
reveal rich details in terms of its dependence on particle type,
transverse ($p_t$) and longitudinal momentum ($\eta$) variables,
and on the centrality and the bombarding energy of the collision.
In the soft transverse momentum region ($p_t \lesssim 2$~GeV/c)
measurements at mid-rapidity are found to be well described by
hydrodynamical
models~\cite{Adcox:2004mh,Adams:2005dq,Csanad:2003qa,Hama:2005dz,Broniowski:2002wp}.
Important is, that in contrast to a uniform distribution of
particles expected in a gas-like system, this liquid behavior
means that the interaction in the medium of these copiously
produced particles is rather strong, as one expects from a fluid.
Detailed investigation of these phenomena suggests that this
liquid flows with almost no viscosity~\cite{Adare:2006ti}.

\begin{figure}
  \begin{center}
  \includegraphics[width=0.8\linewidth]{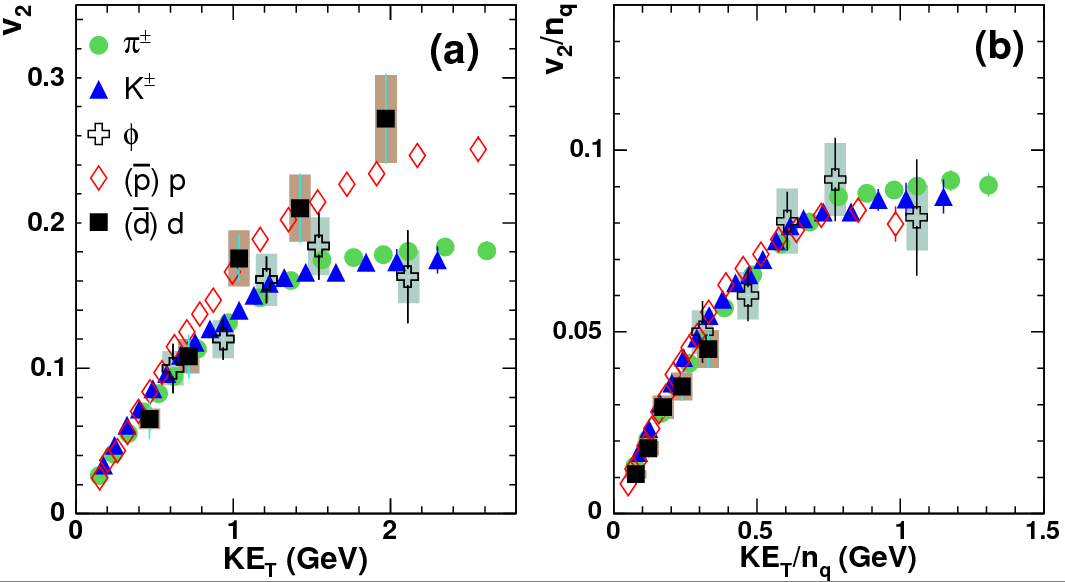}
  \end{center}
  \caption{(color online)(a) $v_2$ vs $KE_T$ for several identified particle
species obtained in mid-central (20-60\%) Au+Au collisions. (b)
$v_2/n_q$ vs $KE_T/n_q$ for the same particle species shown in
panel (a). The shaded bands indicate systematic error estimates
for $(\overline{d})d$ and $\phi$ mesons (see text).}\label{f:v2sc}
\end{figure}

Measurement of elliptic flow of pions, kaons, protons, $\phi$
mesons and deuterons in Au+Au collisions at $\sqrt{s_{NN}}~=~200$
GeV, when plotted against scaling variable $KE_T$ (transverse
kinetic energy) confirm the prediction of perfect fluid
hydrodynamics, that the relatively ``complicated" dependence of
azimuthal anisotropy on transverse momentum and particle type can
be scaled to a single
function\cite{Adare:2006ti,Csanad:2005gv,Csanad:2006sp,Borghini:2005kd,Bhalerao:2005mm}.
On the left plot of Fig.~\ref{f:v2sc} we show this scaling. Mesons
and barions gather into two different groups here. If one scales
both axes of these plots by the number of constituent quarks of
the measured hadrons (as shown on the right plot of
Fig.~\ref{f:v2sc}), the two curves collapse to
one~\cite{Afanasiev:2007tv}. Thus it appears that quark
collectivity dominates the expansion dynamics of these collisions-

\section{Heavy flavour}

\begin{figure}
  \begin{center}
  \includegraphics[width=0.8\linewidth]{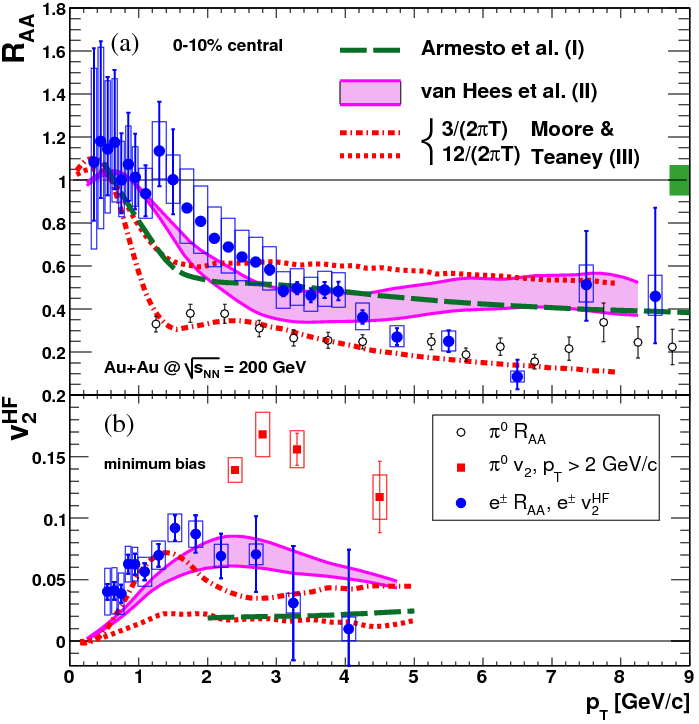}
  \end{center}
  \caption{
(a) $R_{\rm AA}$ of heavy-flavor electrons in 0-10\% central
collisions compared with $\pi^0$ data~\cite{Adler:2003qi} and
model calculations (curves I~\cite{Armesto:2005mz},
II~\cite{vanHees:2005wb}, and III~\cite{Moore:2004tg}). (b)
$v_2^{\rm HF}$ of heavy-flavor electrons in minimum bias
collisions compared with $\pi^0$ data~\cite{Adler:2005rg} and the
same models. Boxes show systematic uncertainty in both
plots.}\label{f:heavyfl}
\end{figure}

We also have measured electrons from heavy flavor (charm and
bottom) decays in Au+Au collisions at $\sqrt{s_{\rm NN}}$ = 200
GeV. The nuclear modification factor $R_{\rm AA}$ relative to p+p
collisions shows a strong suppression in central Au+Au collisions,
indicating substantial energy loss of heavy quarks in the medium
produced at RHIC energies. A large elliptic flow, $v_2$ is also
observed indicating substantial heavy flavor elliptic flow. Both
$R_{\rm AA}$ and $v_2$ show a $p_t$ dependence different from
those of neutral pions. A comparison to transport models which
simultaneously describe $R_{\rm AA}(p_t)$ and $v_2(p_t)$ suggests
that the viscosity to entropy density ratio is close to the
conjectured quantum lower bound, {\it i.e.} near a perfect
fluid~\cite{Armesto:2005mz,vanHees:2005wb,Moore:2004tg}, as shown
on Fig.~\ref{f:heavyfl}

We see, that even heavy flavour is suppressed beyond
extrapolations from cold nuclear matter effects, and even heavy
flavour is flowing similarly to hadrons made out of light quarks.
This suggests strong coupling of charm and bottom to the
medium~\cite{Adler:2003ii,Adare:2006ns}.

\section{Chiral dynamics}

Correlation functions are important to see the collective
properties of particles and the space-time structure of the
emitting source, e.g.\ the observed size of a system can be
measured by two-particle Bose-Einstein
correlations~\cite{HanburyBrown:1956pf}.

The $m_t$ dependence of the strength of the two-pion Bose-Einstein
correlation function $\lambda$ can be used to extract information
on the mass-reduction of the $\eta$' meson (the ninth, would-be
Goldstone-boson), a signal of the U$_{\rm A}(1)$ symmetry
restoration in hot and dense matter: It is known, that if the
chiral U$_{\rm A}$(1) symmetry is restored, then the mass of the
$\eta'$ boson is tremendously decreasing and its production cross
section tremendously increasing. Thus $\eta'$ bosons are copiously
produced, and decaying through $\eta$ bosons (with a very long
lifetime) into low momentum pions. Hence the strength of the
two-particle correlation functions at low relative momenta might
change
significantly.~\cite{Vance:1998wd,Kapusta:1995ww,Huang:1995fc,Hatsuda:1994pi}.

\begin{figure}
  \begin{center}
  \includegraphics[width=0.6\linewidth]{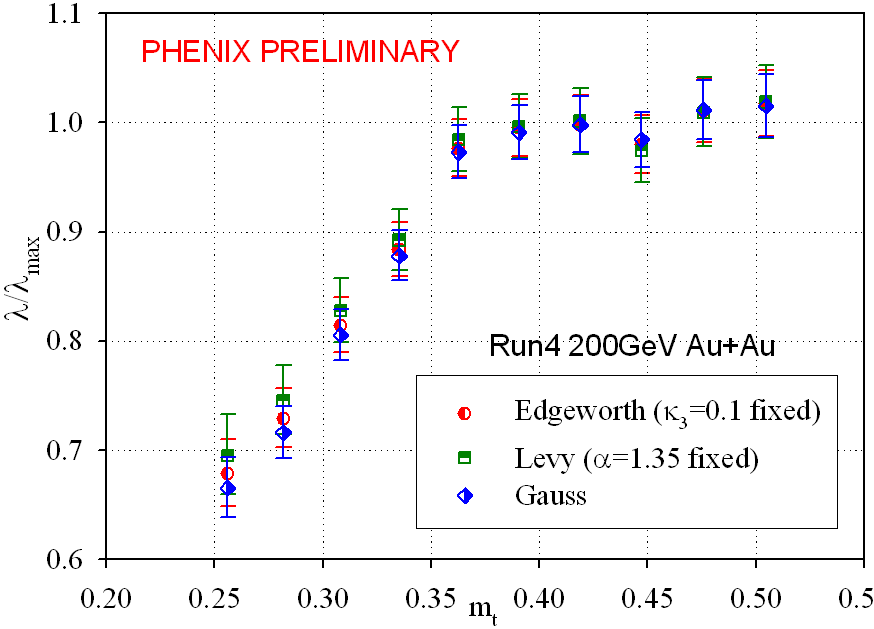}
  \end{center}
  \caption{Measured $\lambda(m_t)$ from different methods}\label{f:ua1}
\end{figure}

PHENIX analyzed~\cite{Csanad:2005nr} two-pion Bose-Einstein
correlations with fits to two-pion correlation functions using
three different shapes, Gauss, Levy and Edgeworth, and determined
$\lambda(m_t)$ from it, as described in
refs.~\cite{Csorgo:1999sj,Csanad:2005nr,Csorgo:2003uv}. We
re-normed the $\lambda(m_t)$ curves with their maximal value on
the investigated $m_t$ interval. This way they all show the same
shape, as shown in Fig.~\ref{f:ua1}. This confirms the existence
and characteristics of the hole in the $\lambda(m_t)$
distribution.

We conclude that at present, results are critically dependent on
our understanding of statistical and systematic errors, and
additional analysis is required to make a definitive statement.

\begin{figure}
  \begin{center}
  \includegraphics[width=0.49\linewidth]{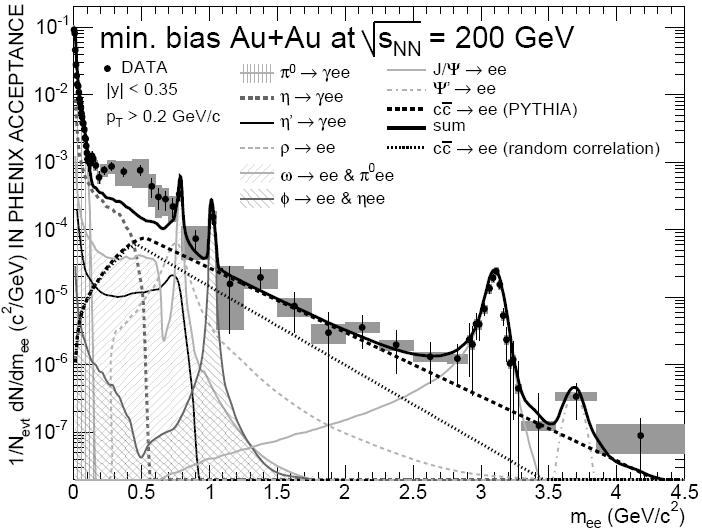}
  \includegraphics[width=0.49\linewidth]{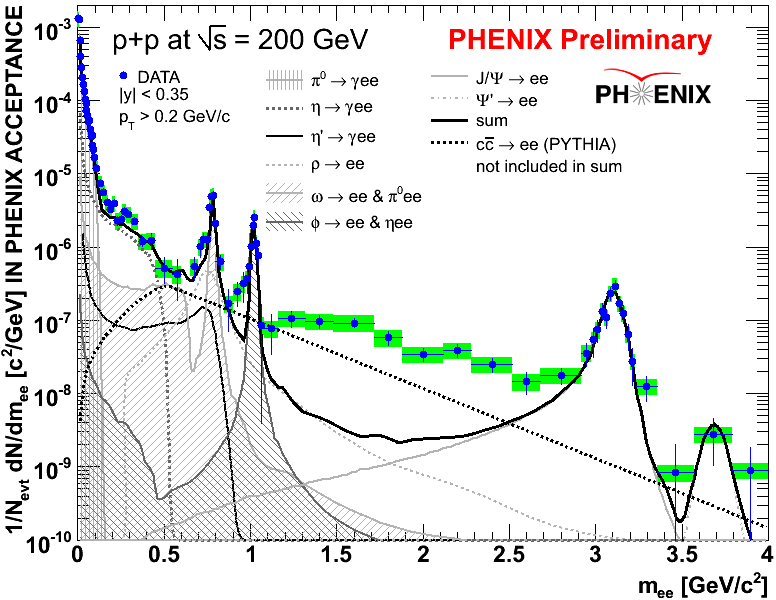}
  \end{center}
  \caption{
Invariant e$^+$e$^-$--pair yield of
refs.~\cite{Afanasiev:2007xw,Toia:2006zh} compared to the yield
from the model of hadron decays. The charmed meson decay
contribution based on PYTHIA is included in the sum of sources
(solid black line). The charm contribution expected if the dynamic
correlation of $c$ and $\bar{c}$ is removed is shown separately.
Statistical (bars) and systematic (boxes) uncertainties are shown
separately; the mass range covered by each data point is given by
horizontal bars. The systematic uncertainty on the cocktail is not
shown.}\label{f:diel}
\end{figure}

The PHENIX experiment has also measured the dielectron continuum
in $\sqrt{s_{NN}}$=200 GeV Au+Au
collisions~\cite{Afanasiev:2007xw,Toia:2006zh}. The data below 150
MeV/c$^2$ are well described by the cocktail of hadronic sources.
The vector mesons $\omega$, $\phi$ and $J/\psi$ are reproduced
within the uncertainties. However, in minimum bias collisions, the
yield is substantially enhanced above the expected yield in the
continuum region from 150 to 750 MeV/c$^2$. The enhancement in
this mass range is a factor of 3.4 $\pm$ 0.2(stat.) $\pm$
1.3(syst.) $\pm$ 0.7(model), where the first error is the
statistical error, the second the systematic uncertainty of the
data, and the last error is an estimate of the uncertainty of the
expected yield. Above the $\phi$ meson mass the data seem to be
well described by the continuum calculation based on PYTHIA, as
shown in Fig.~\ref{f:diel}

\section{Summary and conclusions}

Based on the measurements of suppression of high transverse
momentum hadrons and of their elliptic flow, we can make the
definitive statement, that in relativistic Au+Au collisions
observed at RHIC we see a strongly interacting matter, that has
the characteristics of a perfect fluid. We also see signals of
chiral dynamics by the enhancement of the dielectron continuum
above the expected yield from hadron production and the possible
mass modification of the $\eta$' meson. Future plan is to explore
all properties of the Quark Matter, by analyzing more data and
using higher luminosity.

\bibliographystyle{prlsty}
\bibliography{Master}

\end{document}